\documentclass[runningheads]{llncs}
\usepackage{graphicx}
\usepackage{float}
\usepackage{subcaption}
\usepackage{url}
\usepackage{amsmath}
\usepackage{multirow}
\usepackage{listings}
\graphicspath{{figures/}}

\begin{document}
%
\title{Emotion4MIDI: a Lyrics-based Emotion-Labeled Symbolic Music Dataset}

\author{Serkan Sulun\inst{1,2} \and
Pedro Oliveira\inst{2} \and
Paula Viana\inst{1,3}}
\authorrunning{S. Sulun et al.}

\institute{Institute for Systems and Computer Engineering, Technology and Science (INESC TEC), Porto, Portugal\\
\email{\{serkan.sulun,paula.viana\}@inesctec.pt}
\and
Faculty of Engineering, University of Porto, Porto, Portugal
\email{up201707038@edu.fe.up.pt}\\
\and
ISEP, Polytechnic of Porto, School of Engineering, Porto, Portugal
\email{pmv@isep.ipp.pt}}

\maketitle              
\begin{abstract}
We present a new large-scale emotion-labeled symbolic music dataset consisting of $12k$ MIDI songs. To create this dataset, we first trained emotion classification models on the GoEmotions dataset, achieving state-of-the-art results with a model half the size of the baseline. We then applied these models to lyrics from two large-scale MIDI datasets. Our dataset covers a wide range of fine-grained emotions, providing a valuable resource to explore the connection between music and emotions and, especially, to develop models that can generate music based on specific emotions. Our code for inference, trained models, and datasets are available online.

\end{abstract}

\section{Introduction}
\label{sec:intro}
Music has long been a powerful medium for emotional expression and communication \cite{music_emotion}. The emotional response that music elicits has been studied by scholars from various fields such as psychology \cite{psychology}, musicology \cite{musicology}, and neuroscience \cite{neuroscience}. Especially with the advent of deep learning, there has been an increasing interest in developing machine learning algorithms to automatically analyze and generate music that can evoke specific emotions in listeners \cite{music_survey}. 

Symbolic music -- or MIDI (Musical Instrument Digital Interface) as it is used interchangeably -- is represented as a sequence of notes and is a popular choice for machine learning models due to its compact and structured representation. Large raw MIDI datasets \cite{lakhdataset,redditmididataset} enable unsupervised training of deep neural networks to automatically generate symbolic music. Similar to language modeling, these networks learn to predict the next token i.e. the next note, and at inference time, generate output autoregressively, one token at a time.

However, a human composer's creative process does not simply involve mechanically writing one note after another; it often includes high-level concepts such as motifs, themes and, ultimately, emotions \cite{composing_w_emotion}. To train deep neural networks to generate music based on emotions, large datasets of symbolic music annotated with emotional labels are required. Although there are some publicly available datasets with emotional labels, they are relatively small and do not cover a wide range of emotional states \cite{sulun}.

To address this issue, we present a new large-scale emotion-labeled symbolic music dataset created by analyzing the lyrics of the songs. Our approach leverages the natural connection between lyrics and music, established through emotions. To this end, we first trained models for emotion classification from text on GoEmotions \cite{goemotions}, one of the largest text datasets with $28$ fine-grained emotion labels. Using a model that is half the size of the baseline model, we obtained state-of-the-art results on this dataset. Later, we applied this model to the lyrics of songs from two of the biggest available MIDI datasets, namely Lakh MIDI dataset \cite{lakhdataset} and Reddit MIDI dataset \cite{redditmididataset}. Ultimately, we created a symbolic music dataset consisting of $12k$ MIDI songs labeled with fine-grained emotions. We hope that this dataset will encourage further research in the field of affective algorithmic composition and contribute to the development of intelligent music systems that can understand and evoke specific emotions in listeners.

The remaining of this paper has the following structure: after having introduced our aim and the overall results in Section \ref{sec:intro}, Section \ref{sec:relatedwork} presents the current state of the art on the most relevant topics for this work, namely text emotion classification and the existing emotion-labeled symbolic music datasets. Section \ref{sec:methodology} will delve into the proposed solution describing all the implemented steps, while results are presented and discussed in Section \ref{sec:results}. Finally, we conclude by pointing out some possible future work in Section \ref{sec:conclusion}.

\section{Related work}
\label{sec:relatedwork}

\subsection{Text emotion classification}

Emotion classification from text -- or sentiment analysis, as used interchangeably in the machine learning literature -- allows us to automatically identify and/or quantify the emotion expressed in a piece of text, such as a review, social media post, or customer feedback \cite{sentiment_survey}. Identifying the underlying emotion in text is useful in various fields such as customer service \cite{customer}, finance \cite{finance}, politics \cite{politics}, and entertainment \cite{emotionpv}.

Machine learning methods have significantly advanced the state of the art in text emotion classification for the past two decades. However, the earliest works in this field relied on hand-crafted features, such as frequently used n-grams \cite{ngrams}, or adjectives and adverbs that are associated with particular emotions \cite{appraisal}. Nonetheless, the advent of deep learning has made it computationally feasible to process raw inputs without extracting features manually, leading to better performance \cite{alexnet}. Recurrent Neural Networks and their improved variants such as Long Short-Term Memory were initially used \cite{rnn} but were later replaced by the transformer model \cite{transformer}, which is the current state of the art in natural language processing (NLP) tasks.

Fine-tuning pretrained models on specific tasks has been shown to produce better performance. The GPT (generative pretraining) model is a large transformer that was pretrained on the task of next token prediction and then was fine-tuned on specific NLP tasks, resulting in state-of-the-art performance \cite{gpt}. The BERT (Bidirectional Encoder Representations from Transformers) model improved upon these results by employing masked token prediction as its pretraining task \cite{bert}.

\subsection{Emotion-labeled symbolic music datasets}

MIDI (Musical Instrument Digital Interface) is a symbolic music format widely used to represent musical performances and compositions in the digital domain. MIDI files contain only the musical information, such as the notes, tempo, and dynamics, without the sound itself, like a ``digital music sheet". Compared to audio formats, MIDI files have a smaller size and dimensionality, which makes them more manageable and suitable for modeling with deep neural networks \cite{music_survey}.

The majority of existing literature on symbolic music generation relies on a non-conditional approach. In other words, these methods are trained on raw MIDI data without any explicit labels, allowing them to generate new music that is similar to the examples in the training dataset \cite{music_transformer}. Some approaches, however, leverage low-level features within the data to create music in a conditional way \cite{counterpoint}. For instance, they might use short melodies, chords, or single-instrument tracks as a basis for generating corresponding melodies. While such methods could be considered as ``conditional", they do not make use of specific labels and are thus unable to capture high-level factors such as emotions or genres.

Using emotion as the specific high-level condition gives 
rise to the field of ``affective algorithmic composition" (AAC) \cite{aac}. However, the development of machine learning AAC models is currently limited by the lack of large-scale symbolic music datasets with emotion labels. Some existing datasets include VGMIDI, which contains $204$ piano-based video game soundtracks with continuous valence and arousal labels \cite{vgmidi}, Panda et al., which includes $193$ samples with discrete emotion labels \cite{mirexlike}, and EMOPIA, which consists of $387$ piano-based pop songs with four emotion labels \cite{emopia}. Unfortunately, due to their small sizes, these datasets are insufficient for training deep neural networks with millions of parameters. 
Sulun et al. addressed this issue by labeling $34k$ samples with continuous valence and arousal labels \cite{sulun}. Though initially designed for audio samples, these labels were matched to their corresponding MIDI files to train emotion-based symbolic music generators that produced output music with emotional coherence. 
While this study exploited the correspondence between audio and symbolic music, there has been no utilization of the correspondence between lyrics and symbolic music to acquire high-level semantic labels.

\section{Methodology}\label{sec:methodology}

This section outlines the steps we followed to achieve our goal of creating a symbolic music dataset with emotion labels. Specifically, we begin by describing the model utilized for emotion classification, followed by a discussion of the training process, and conclude with an overview of how the model was applied to song lyrics to extract the corresponding emotion labels.

\subsection{Model}

We employ DistilBERT as the backbone of our model \cite{distilbert}, which is a condensed and compressed variant of the BERT (Bidirectional Encoder Representations from Transformers) model \cite{bert}, achieved through knowledge distillation \cite{distillation_2006,distillation_2015}. DistilBERT utilizes fewer layers than BERT and learns from BERT's outputs to mimic its behavior. Our model consists of $6$ layers, with each layer containing $12$ attention heads and a dimensionality of $768$, yielding a total of $67M$ parameters. To facilitate multi-label classification, we have customized the output layer while adding a sigmoid activation layer at the end. The output layer's size is determined by the number of labels present in the training dataset, which can be either $7$ or $28$.

\subsection{Training}
The first step towards our aim of building an emotion-labeled symbolic music dataset is to train the model to perform multi-label emotion classification based on text input. 

\subsubsection{Dataset}

We trained our model using the GoEmotions dataset \cite{goemotions}. This dataset consists of English comments from the website \textit{reddit.com}, which were manually annotated to identify the underlying emotions. It is a multi-label dataset, which means that each comment can have more than one emotion label. The dataset comprises $27$ emotions and a ``neutral" label. The labels are further grouped into $7$ categories, including the six basic emotions identified by Ekman (joy, anger, fear, sadness, disgust, and surprise) as well as the ``neutral" label \cite{ekman}. The dataset has a total of $58k$ samples, which were split into training, validation, and testing sets in the ratio of $80\%$, $10\%$, and $10\%$, respectively. Given the number of labels and its size, GoEmotions is one of the largest emotion classification datasets and has the highest number of discrete emotion labels \cite{emotion_review}.

\subsubsection{Training and evaluation metrics}

We trained our models using binary cross-entropy loss. For evaluation, we used precision, recall, and F1-score, with macro averaging. The decision cutoff was set at $0.3$, meaning that predictions with a value of $0.3$ or greater are considered positive predictions and others negative.

\subsubsection{Implementation details}

We trained two models to classify a given text into $7$ and $28$ labels. We used a dropout rate of $0.1$ and a gradient clipping norm of $1$. The batch size was set to $16$ for the model with $7$ output labels and to $32$ for the model with $28$ output labels. We applied a learning rate of $5e-5$ for the former and $3e-5$ for the latter. We used early stopping considering the F1-score on the validation dataset, which corresponded to training for $10$ epochs for both models. We implemented the models using Huggingface library \cite{huggingface} with Pytorch backend \cite{pytorch} and trained them using a single Nvidia GeForce GTX 1080 Ti GPU.

\subsection{Inference}

After training the model for text-based emotion classification, we used it in inference mode, using the song lyrics from the MIDI files as inputs. This allowed us to create a MIDI dataset labeled with emotions.

\subsubsection{Datasets}

We used two MIDI datasets that are publicly available and were created by gathering MIDI files from various online sources: 
the Lakh MIDI dataset consisting of $176k$ samples \cite{lakhdataset} and the Reddit MIDI dataset containing $130k$ samples \cite{redditmididataset}. We filtered the datasets by selecting MIDI files that contain lyrics in the English language with at least $50$ words. This filtering process resulted in a total of $12509$ files, consisting of $8386$ files from the Lakh MIDI dataset and $4123$ files from the Reddit MIDI dataset. During inference, we utilized the two pretrained models, feeding the entire song's lyrics, using a truncation length of $512$.

\section{Results}
\label{sec:results}
In this section, we will first present the emotion classification performance of our trained models. Then, we will introduce the emotion-labeled MIDI dataset, which we created by analyzing the sentiment of the song lyrics using our trained models.

\subsection{Emotion classification on the GoEmotions dataset}

We evaluated the performance of our trained models on the test split of the GoEmotions dataset and compared our results with the baseline presented in the original paper \cite{goemotions}. Similar to the original paper, we report our results for scenarios using two sets of labels, with $7$ and $28$ emotions. For each label, we reported the precision, recall, and F1-scores along with the macro-averages. It is important to mention that, as the dataset is imbalanced, macro-averaging is more appropriate than micro-averaging, as it was also used in the original paper. We note that the baseline model is BERT and has twice the size of our model \cite{bert}. 

The trade-off between precision and recall is determined by the cutoff value. Therefore, we emphasize higher F1-scores because they provide a more balanced perspective by taking the harmonic mean of precision and recall, and are much less sensitive to the cutoff value. Although the original paper did not state the cutoff value, we achieved the best F1-score and similar performance to the original paper on the $7$-label dataset using a cutoff value of $0.3$. For consistency, we used the same value for the $28$-label dataset. We present our results on the dataset with $7$ and $28$ labels in Tables \ref{table:7labels} and \ref{table:28labels}, respectively.

\begin{table}[th]
  \scriptsize
  \centering
  \caption{7-label classification results}
    \begin{tabular}{lcc|cc|cc}
    \\
          & \multicolumn{2}{c}{Precision} & \multicolumn{2}{c}{Recall} & \multicolumn{2}{c}{F1-score} \\
          \hline 
          & Baseline & Ours & Baseline & Ours & Baseline & Ours \\
          \hline
    anger & 0.50  & 0.50  & 0.65  & 0.67  & \textbf{0.57}  & \textbf{0.57} \\
    disgust & 0.52  & 0.57  & 0.53  & 0.49  & \textbf{0.53}  & 0.52 \\
    fear  & 0.61  & 0.57  & 0.76  & 0.73  & \textbf{0.68}  & 0.64 \\
    joy   & 0.77  & 0.75  & 0.88  & 0.89  & \textbf{0.82}  & \textbf{0.82} \\
    neutral & 0.66  & 0.63  & 0.67  & 0.75  & 0.66  & \textbf{0.68} \\
    sadness & 0.56  & 0.57  & 0.62  & 0.67  & 0.59  & \textbf{0.61} \\
    surprise & 0.53  & 0.59  & 0.70  & 0.62  & \textbf{0.61}  & \textbf{0.61}  \\ \hline

    macro-average & 0.59  & 0.60 & 0.69  & 0.69  & \textbf{0.64}  & \textbf{0.64} \\

    \end{tabular}
  \label{table:7labels}
\end{table}

\begin{table}[H]
  \scriptsize
  \centering
  \caption{28-label classification results}
\begin{tabular}{lcc|cc|cc}
\\
               & \multicolumn{2}{c}{Precision} & \multicolumn{2}{c}{Recall} & \multicolumn{2}{c}{F1-score}        \\ \hline
               & Baseline        & Ours        & Baseline       & Ours      & Baseline      & Ours          \\ \hline
admiration     & 0.53            & 0.65        & 0.83           & 0.75      & 0.65          & \textbf{0.70} \\
amusement      & 0.70            & 0.72        & 0.94           & 0.91      & 0.80          & \textbf{0.81} \\
anger          & 0.36            & 0.53        & 0.66           & 0.49      & 0.47          & \textbf{0.51} \\
annoyance      & 0.24            & 0.40        & 0.63           & 0.31      & 0.34          & \textbf{0.35} \\
approval       & 0.26            & 0.39        & 0.57           & 0.38      & 0.36          & \textbf{0.39} \\
caring         & 0.30            & 0.37        & 0.56           & 0.46      & 0.39          & \textbf{0.41} \\
confusion      & 0.24            & 0.52        & 0.76           & 0.42      & 0.37          & \textbf{0.47} \\
curiosity      & 0.40            & 0.47        & 0.84           & 0.62      & \textbf{0.54} & 0.53          \\
desire         & 0.43            & 0.66        & 0.59           & 0.42      & 0.49          & \textbf{0.51} \\
disappointment & 0.19            & 0.39        & 0.52           & 0.22      & \textbf{0.28} & \textbf{0.28} \\
disapproval    & 0.29            & 0.39        & 0.61           & 0.41      & 0.39          & \textbf{0.40} \\
disgust        & 0.34            & 0.64        & 0.66           & 0.39      & 0.45          & \textbf{0.48} \\
embarrassment  & 0.39            & 0.72        & 0.49           & 0.35      & 0.43          & \textbf{0.47} \\
excitement     & 0.26            & 0.43        & 0.52           & 0.47      & 0.34          & \textbf{0.45} \\
fear           & 0.46            & 0.60        & 0.85           & 0.76      & 0.60          & \textbf{0.67} \\
gratitude      & 0.79            & 0.88        & 0.95           & 0.92      & 0.86          & \textbf{0.90} \\
grief          & 0.00            & 0.00        & 0.00           & 0.00      & \textbf{0.00} & \textbf{0.00} \\
joy            & 0.39            & 0.59        & 0.73           & 0.61      & 0.51          & \textbf{0.60} \\
love           & 0.68            & 0.78        & 0.92           & 0.85      & 0.78          & \textbf{0.81} \\
nervousness    & 0.28            & 0.45        & 0.48           & 0.43      & 0.35          & \textbf{0.44} \\
neutral        & 0.56            & 0.61        & 0.84           & 0.76      & \textbf{0.68} & \textbf{0.68} \\
optimism       & 0.41            & 0.56        & 0.69           & 0.52      & 0.51          & \textbf{0.54} \\
pride          & 0.67            & 0.83        & 0.25           & 0.31      & 0.36          & \textbf{0.45} \\
realization    & 0.16            & 0.39        & 0.29           & 0.14      & \textbf{0.21} & \textbf{0.21} \\
relief         & 0.50            & 0.00        & 0.09           & 0.00      & \textbf{0.15} & 0.00          \\
remorse        & 0.53            & 0.59        & 0.88           & 0.86      & 0.66          & \textbf{0.70} \\
sadness        & 0.38            & 0.57        & 0.71           & 0.60      & 0.49          & \textbf{0.59} \\
surprise       & 0.40            & 0.56        & 0.66           & 0.50      & 0.50          & \textbf{0.53} \\ \hline
macro-average  & 0.40            & 0.53        & 0.63           & 0.50      & 0.46          & \textbf{0.50}
\end{tabular}
\label{table:28labels}
\end{table}

Based on the F1-scores, our model performs comparably to the baseline on the $7$-label dataset. Specifically, our model has a better performance on $2$ labels, worse on $2$ labels, and the same on $3$ labels, as well as for the macro-average. On the $28$-label dataset, our model surpasses the baseline with only a lower performance on $2$ labels, equal performance on $4$ labels, and better performance on the remaining $22$ labels. Furthermore, our model demonstrates an improvement of $0.04$ in terms of the macro-average.

We hypothesize that a smaller model, such as ours (DistilBERT), may perform better than a larger baseline model (BERT) in certain settings, such as when there are a limited number of training samples or a high output/target dimensionality, as in the case of the $28$-label dataset. In these scenarios, models are more prone to overfitting, as has been previously observed \cite{overfitting}. Additionally, the original paper \cite{distilbert} demonstrates that the DistilBERT model outperforms BERT on the Winograd Natural Language Inference (WNLI) dataset \cite{wnli}.

\subsection{Labeled MIDI dataset}

We used our trained models to analyze the song lyrics of the Lakh and Reddit MIDI datasets, resulting in an augmented dataset that contains the file paths to $12509$ MIDI files and their corresponding predicted probabilities for emotion labels. To provide more flexibility to the users, we did not apply a threshold to the predicted probabilities, allowing the entire dataset to be used as is. We generated two CSV (comma-separated values) files containing the $7$ and $28$ emotion labels as columns, with the $12509$ MIDI file paths as rows. Our code for inference, trained models, and datasets are available online.\footnote{\url{https://github.com/serkansulun/lyricsemotions}}

For demonstration purposes, we provide transposed versions of the tables, using only $3$ samples, shown in Tables \ref{table:sample_7labels} and \ref{table:sample_28labels}. We note that the values do not necessarily add up to one, due to the nature of multi-label classification.

\vspace{-4mm}

\begin{table}[th]
\centering
\scriptsize
\caption{Sample entries from the $7$-label dataset. Due to space limitations, the file paths are replaced with the artist and song names and are as the following: John Lennon - Imagine: ``\texttt{lakh/5/58c076b72d5115486c09a7d9e6df1029.mid}" (artist and title obtained using Million Song Dataset \cite{msd}), ABBA - Take a Chance on Me: ``\texttt{reddit/A/ABBA.Take a chance on me K.mid}", Elvis Presley - Are You Lonesome Tonight: ``\texttt{reddit/P/PRESLEY.Are you lonesome tonight K.mid}"\\}
\label{table:sample_7labels}

\begin{tabular}{l@{\hskip 6mm}c@{\hskip 9mm}c@{\hskip 3mm}c}
           & \begin{tabular}{@{}c@{}}John Lennon \\ - Imagine \end{tabular} & \begin{tabular}{@{}c@{}}ABBA - Take \\ a Chance on Me\end{tabular} & \begin{tabular}{@{}c@{}}Elvis Presley - Are \\ You Lonesome Tonight\end{tabular}  \\ \hline
anger    & 0.0051                                      & 0.0146                                  & 0.0272                                          \\
disgust  & 0.0003                                      & 0.0009                                  & 0.0045                                          \\
fear     & 0.0005                                      & 0.0024                                  & 0.0131                                          \\
joy      & \textbf{0.8072}                                      & \textbf{0.8948}                                  & 0.0477                                          \\
neutral  & 0.1953                                      & 0.1420                                  & 0.0782                                          \\
sadness  & 0.0013                                      & 0.0069                                  & \textbf{0.7372}                                          \\
surprise & 0.0754                                      & 0.0053                                  & 0.5465                                         
\end{tabular}
\end{table}

\begin{table}[h]
\vspace{-3mm}
\centering
\scriptsize
\caption{Sample entries from the $28$-label dataset.} 
\vspace{3mm}
\label{table:sample_28labels}

\begin{tabular}{l@{\hskip 6mm}c@{\hskip 9mm}c@{\hskip 3mm}c}
           & \begin{tabular}{@{}c@{}}John Lennon \\ - Imagine \end{tabular} & \begin{tabular}{@{}c@{}}ABBA - Take \\ a Chance on Me\end{tabular} & \begin{tabular}{@{}c@{}}Elvis Presley - Are \\ You Lonesome Tonight\end{tabular}  \\ \hline
admiration     & 0.0021                                      & 0.0091                                  & 0.0048                                          \\
amusement      & 0.0051                                      & 0.0012                                  & 0.0027                                          \\
anger          & 0.0025                                      & 0.0018                                  & 0.0053                                          \\
annoyance      & 0.0024                                      & 0.0020                                   & 0.0075                                          \\
approval       & 0.0026                                      & 0.0809                                  & 0.0072                                          \\
caring         & 0.0067                                      & \textbf{0.6169}                                  & 0.0601                                          \\
confusion      & 0.0070                                       & 0.0035                                  & 0.1029                                          \\
curiosity      & 0.0332                                      & 0.0141                                  & \textbf{0.6502}                                          \\
desire         & 0.0482                                      & 0.0472                                  & 0.0055                                          \\
disappointment & 0.0044                                      & 0.0016                                  & 0.0199                                          \\
disapproval    & 0.0019                                      & 0.0030                                   & 0.0048                                          \\
disgust        & 0.0007                                      & 0.0003                                  & 0.0009                                          \\
embarrassment  & 0.0006                                      & 0.0002                                  & 0.0045                                          \\
excitement     & 0.0130                                       & 0.0049                                  & 0.0011                                          \\
fear           & 0.0026                                      & 0.0026                                  & 0.0035                                          \\
gratitude      & 0.0007                                      & 0.0017                                  & 0.0059                                          \\
grief          & 0.0008                                      & 0.0016                                  & 0.0085                                          \\
joy            & 0.0025                                      & 0.0040                                   & 0.0018                                          \\
love           & 0.0021                                      & 0.1079                                  & 0.0193                                          \\
nervousness    & 0.0007                                      & 0.0017                                  & 0.0094                                          \\
neutral        & 0.2954                                      & 0.4288                                  & 0.0757                                          \\
optimism       & \textbf{0.7554}                                      & 0.1423                                  & 0.0060                                           \\
pride          & 0.0010                                       & 0.0013                                  & 0.0006                                          \\
realization    & 0.0023                                      & 0.0040                                   & 0.0045                                          \\
relief         & 0.0004                                      & 0.0033                                  & 0.0011                                          \\
remorse        & 0.0005                                      & 0.0012                                  & 0.1491                                          \\
sadness        & 0.0011                                      & 0.0027                                  & 0.1767                                          \\
surprise       & 0.0107                                      & 0.0005                                  & 0.0020                                          
\end{tabular}

\end{table}

For further demonstration and ease of analysis, we provide excerpts from the lyrics of each of the three sample songs in Listing \ref{lyrics}, along with the emotions having predicted probabilities higher than $0.1$ in descending order. It is noteworthy that having a dataset with $28$ emotion labels allows for a more nuanced representation of emotions. For instance, when we examine this dataset, the song ``Imagine" is predicted to have ``optimism" as its top emotion, whereas ``Take a Chance on Me" is predicted to have ``caring" as its top emotion. However, both songs are predicted to have ``joy" as their top emotion in the dataset with only seven labels.

\begin{figure}[ht]
\includegraphics[width=\textwidth]{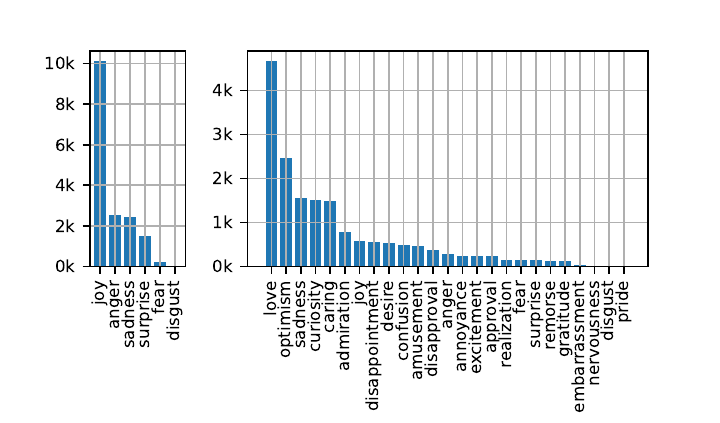}
\caption{The number of samples containing each emotion in our $7$-label (left) and $28$-label (right) datasets. The ``neutral" label is excluded. Emotions with a prediction value higher than $0.1$ are considered positive labels.} 
\label{fig:counts}
\end{figure}

We also present the number of samples containing each emotion in our datasets in Figure \ref{fig:counts}. In these figures, we excluded the ``neutral" label and considered emotions with a prediction value higher than $0.1$ as positive labels. 

\begin{lstlisting}[float,floatplacement=h,basicstyle=\ttfamily\fontsize{7.0}{10}\selectfont,caption={Sample entries with excerpts from lyrics, and emotions with a predicted value higher than 0.1.},label=lyrics]
File path: lakh/5/58c076b72d5115486c09a7d9e6df1029.mid
Artist - Title: John Lennon - Imagine
Lyrics: 
    Imagine there's no heaven. 
    It's easy if you try. 
    No hell below us. 
    Above us, only sky. 
    Imagine all the people. 
    Livin' for today.
7-label predictions:
    joy:            0.8072
    neutral:        0.1953
28-label predictions:
    optimism:       0.7554
    neutral:        0.2954

File path: reddit/A/ABBA.Take a chance on me K.mid
Artist - Title: ABBA - Take a Chance on Me
Lyrics: 
    If you change your mind, I'm the first in line.
    Honey, I'm still free. 
    Take a chance on me.
    If you need me, let me know, gonna be around.
    If you've got no place to go, if you're feeling down.
7-label predictions:
    joy:            0.8948
    neutral:        0.1420
28-label predictions:
    caring:         0.6169
    neutral:        0.4288
    optimism:       0.1423
    love:           0.1079

File path: reddit/P/PRESLEY.Are you lonesome tonight K.mid
Artist - Title: Elvis Presley - Are You Lonesome Tonight
Lyrics: 
    Are you lonesome tonight? 
    Do you miss me tonight? 
    Are you sorry we drifted apart? 
    Does your memory stray to a bright summer day, 
    When I kissed you and called you sweetheart?
7-label predictions:
    sadness:        0.7372
    surprise:       0.5465
28-label predictions:
    curiosity:      0.6502
    sadness:        0.1767
    remorse:        0.1491    
    confusion:      0.1029
\end{lstlisting}

\section{Conclusion and future work}
\label{sec:conclusion}

In this work, we first trained models on the largest text-based emotion classification dataset, GoEmotions, in both $7$-label and $28$-label variants \cite{goemotions}. We achieved state-of-the-art results using a model half the size of the baseline. We then used these trained models to analyze the emotions of the song lyrics from the two largest MIDI datasets, Lakh MIDI dataset \cite{lakhdataset} and Reddit MIDI dataset \cite{redditmididataset}.  This analysis resulted in an augmented dataset of $12509$ MIDI files with emotion labels in a multi-label format, using either $7$ basic-level or $28$ fine-grained emotions. We made the datasets, inference code, and trained models available for researchers to use in various tasks, including symbolic music processing, natural language processing, and sentiment analysis. 

In our future work, we plan to further narrow the considerable gap between symbolic music and emotion. In particular, we aim to create superior models that can automatically compose music that is based on emotions or user-provided input. We believe that incorporating emotions is vital in composing music, hence it can help to push the boundaries of computational creativity, bringing it one step closer to human-like performance.

\bibliographystyle{splncs04}
\bibliography{bibliography}

\end{document}